\documentclass[twocolumn,amsmath,amssymb,prl,aps,floatfix,showpacs]{revtex4}
\usepackage{graphicx}
\usepackage{bm}
\usepackage{amsmath,amsfonts,color}

\begin{document}


\title{Spin Josephson vortices in two tunnel coupled spinor Bose gases}

\author{T.W.A. Montgomery}
\author{W. Li}
\author{T.M. Fromhold}

\affiliation{School of Physics and Astronomy, University of Nottingham, Nottingham, NG7 2RD, UK}
\date{\today}

\begin{abstract}
{We study topological excitations in spin-1 Bose-Einstein condensates trapped in an elongated double-
well optical potential. This system hosts a new topological defect, the spin Josephson vortex (SJV), which 
forms due to the competition between the inter-well atomic tunneling and short-range ferromagnetic 
two-body interaction. We identify the spin structure and formation dynamics of the SJV and determine 
the phase diagram of the system. By exploiting the intrinsic stability of the SJV, we propose a dynamical 
method to create SJVs under realistic experimental conditions.}
\end{abstract}


\maketitle
Ultracold spinor atomic gases that exhibit both superfluidity and magnetic order display an abundance of 
rich static and dynamical properties. This has attracted considerable theoretical and experimental study, 
with particular focus on the topological excitations of trapped spinor gases~\cite{ho_98,ohmi_1998,schmal_04,bigelow_05,chang_05,sadler_06}. 
Topological phases of single trapped spinor gases, such as spin vortices~\cite{,sadler_06,mizushima_2002,saito_06,write_09}, knots \cite{kawag_08} and skyrmions~\cite{usama_01,leslie09,choi12}, 
depend critically on the mean-field order-parameter manifold. A remarkable feature of these dynamical excitations is that their size is typically larger
than the underlying spin healing length (SHL)~\cite{ho_98,ohmi_1998}.

When spinor atoms are confined in optical lattice potentials, atomic tunneling between adjacent lattice sites competes with the spin dependent inter-atomic 
interaction. This competition provides a mechanism for the emergence of topological phases, which have been identified and investigated in several studies~
\cite{pu_01,demler_02,pu_02,yip_03,rizzi_05,song_07,batrouni_09,rodriguez_11}. It can also strongly influence the behavior of a simpler system comprising 
atoms confined in double-well (DW) potentials~\cite{bloch2005,Albiez05}, which are analogous to Josephson junctions in solid state devices. Analysis of such 
systems often uses the lowest energy mode approximation~\cite{milburn_97}, which allows the intra-well spatial motion to be mapped as a function of time. 
Even in this limit, spin-dependent population oscillations between the two potential wells have been identified~\cite{you_05,you_07,julia_09,messeguer_11,wzhang_2012}. 
But beyond this limit, it remains unclear whether quantum fluctuations can trigger the formation of extended topological excitations when the size of the individual spinor 
gases in each well exceeds the spin healing length (SHL). 

\begin{figure}
\includegraphics[width=1.0\columnwidth]{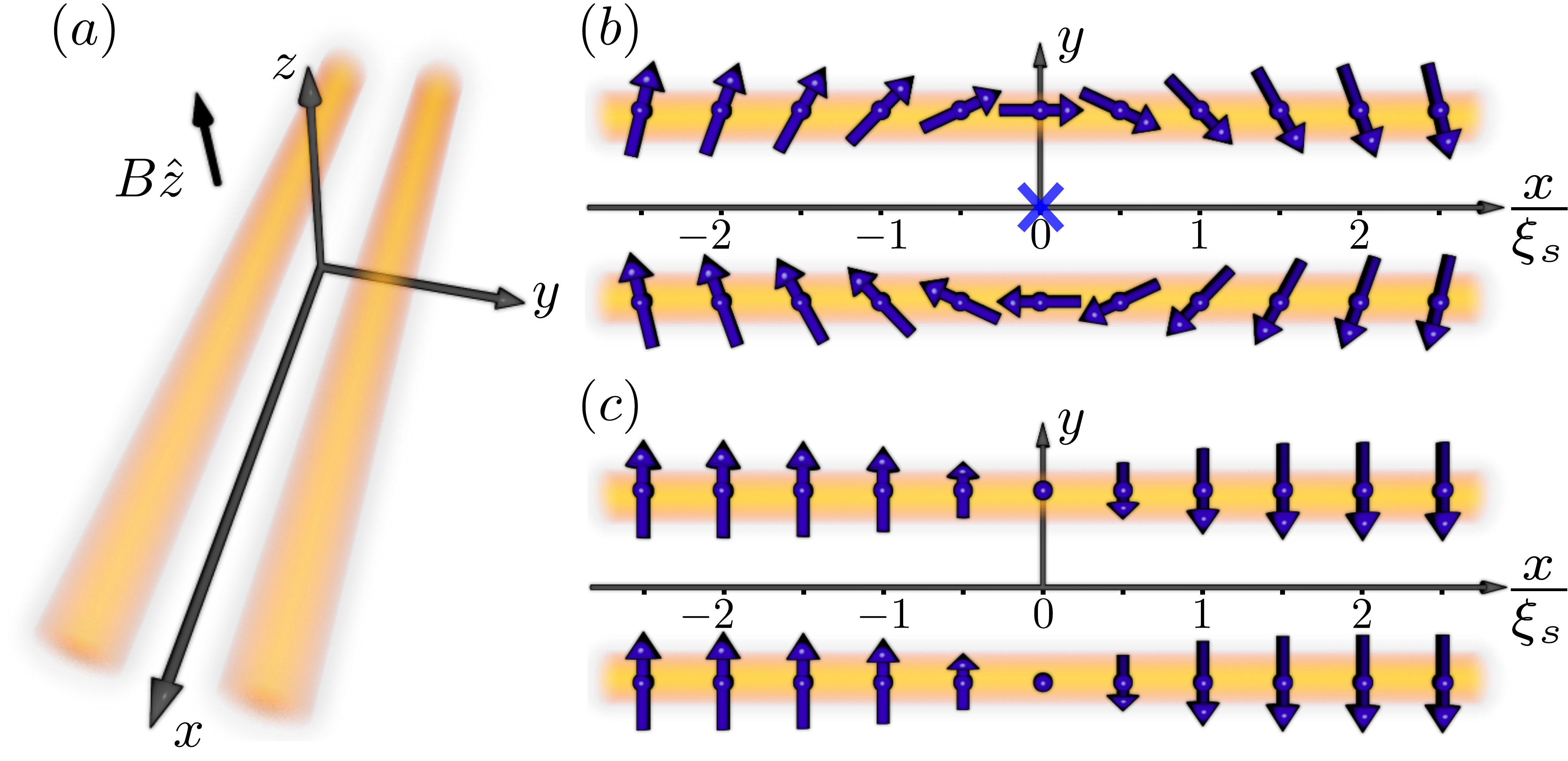}
\caption{(Color online) (a) Schematic diagram of the system. The weakly coupled spin-1 BECs are trapped in a double-well potential, which gives strong (weak) confinement in the 
$y-z$ plane ($x$-direction). A uniform magnetic field $B\hat{z}$ is applied along the $z$-direction. (b,c) show spin vector patterns corresponding to (b) a spin Josephson vortex 
(SJV) centered on the blue cross, (c) a ferromagnetic domain wall (FDW). The parameters are $\kappa = \kappa_{\text{c}}/2$ for the SJV and $\kappa = 2\kappa_{\text{c}}$ for the FDW. 
Other parameters are $q=0$ and $\alpha =\pi/2$. See text for more details of the parameters and spin patterns.} 
\label{fig1}
\end{figure}

In this work, we show that a dynamically stable topological excitation, the so-called spin Josephson vortex (SJV), forms in two weakly coupled spin-1 ferromagnetic Bose-Einstein
condensates (BECs) trapped in an elongated DW potential [Fig.~\ref{fig1}(a)]. As depicted in Fig.~\ref{fig1}(b), a key feature of an SJV is its fixed spin current, facilitated by the 
inter-well atomic tunneling, which circulates about a point mid-way between the two wells. Due to its large size, on the order of several SHLs, the SJV is a macroscopic topological 
object. We determine analytically the parameter space  required for SJVs to form in a uniform system, where they are the only stable topological excitation. We show that, as a 
consequence of this stability, the SJV can be created dynamically through the decay of a ferromagnetic domain wall (FDW)~\cite{saito_052,sadler_06,saito_07}. We demonstrate 
that the SJV can be realized by implementing this dynamical scheme under conditions that can be fully attained with current experimental techniques.

Our system comprises two one-dimensional (1D) spin-1 BECs trapped in a symmetric optical DW potential [Fig.~\ref{fig1}(a)]. Both atom clouds are strongly confined in the transverse ($y,z$) directions. 
The dynamics of weakly coupled spinor BECs may be described by the spin-1 field operator~\cite{zhang05}, ${\bf\hat\Psi}(x) = \hat{\bf\Psi}^l(x) + \hat{\bf\Psi}^r(x)$, where 
$\hat{\bf\Psi}^j(x) = [\hat{\psi}^j_1(x),\hat{\psi}^j_0(x),\hat{\psi}^j_{-1}(x)]^T$ with $j=l \ (r)$ indicating the left (right) well and $m=\{1,0,-1\}$ denoting the three Zeeman levels. A uniform magnetic
field $B{\hat z}$ is applied along the $z$-axis. The many-body Hamiltonian is
\begin{equation}
{\cal H}= {\cal H}_t+ {\cal H}^l+{\cal H}^r,
\label{eq:hamiltonian}
\end{equation}
where ${\cal H}_{t} = -\kappa\int{dx[\hat{\bf\Psi}^{l \dagger}\hat{\bf\Psi}^r+ \text{H.c.}]}$ denotes the inter-well tunneling with state-independent tunneling strength $\kappa$ \cite{milburn_97}. The 
Hamiltonian of atoms in the $j^{\text{th}}$ well is ${\cal H}^j = \int{dx[\hat{\bf\Psi}^{j\dagger}{\bf h}^j \hat{\bf\Psi}^j + \frac{c_0}{2}({\hat n}^j)^2 + \frac{c_1}{2}\vec{\bf F}^j\cdot\vec{\bf F}^j}]$, where 
${\bf h}^j_{m,m'}= -\delta_{m,m'} (\partial^2_x/2-\mu^j-pm+q m^2)$. Here, $\mu^j$, $p=-\mu_{\text{B}}B/2$ and $q=(\mu_{\text{B}}B)^2/4E_{\text{hf}}$ denote the chemical 
potential, linear and quadratic Zeeman energies respectively, where $\mu_{\text{B}}$ is the Bohr magneton and  $E_{\text{hf}}$ is the hyperfine energy splitting~\cite{saito_07}. The two-body collisional 
interactions in the $j^{\text{th}}$ spinor BEC enter the expression for $\mathcal{H}^j$ via the scalar density ${\hat n}^j=\hat{\bf\Psi}^{j\dagger}\hat{\bf\Psi}^j$ and the spin-dependent vector density, 
$\vec{\bf F}^j = \hat{\bf\Psi}^{j\dagger} \vec{\bf f} \ \hat{\bf\Psi}^j$, where $\vec{\bf f}$ is the Cartesian vector of the spin-1 matrices $(f_x,f_y,f_z)$. The effective 1D interaction strengths are 
$c_{0} = 16\hbar^2(a_0+2a_2)/9Mr_{\perp}^2$ and $c_1 = -16\hbar^2(a_0-a_2)/9M r_{\perp}^2$, where $a_S$ is the 3D s-wave scattering length for collisions with total angular momentum 
$S=0,2$~\cite{ho_98,ohmi_1998}, and $r_{\perp}$ is the width of the BEC in the transverse directions.

We study the system using mean field theory. Let us first investigate the stationary state of the system. We use a simple ansatz to describe the order parameter
\begin{equation}
\left(\begin{array}{c}\psi_1^j \\ \psi_0^j \\ \psi_{-1}^j\end{array}\right)=\left(\begin{array}{c} \psi^j_1\\ \sqrt{n^j - 2|\psi^j_1|^2} \\ (\psi^{j}_1)^{*}\end{array}\right),
\end{equation}
where $n^j = {\bf \Psi}^{j\dagger}\mathbf{\Psi}^{j}$ is the total density of the $j^{\text{th}}$ BEC~\cite{saito_07}. For convenience, 
we scale length, time and energy by the variables $\xi_0 = \hbar/(Mc_0 n_R)^{1/2}$, $t_0 = \hbar/c_0 n_R$ and 
$\epsilon_0 =c_0 n_R$, where the reference density, $n_R$, is chosen to ensure correct chemical potentials
in the two atom clouds. Neglecting the spatial dependence of $n^j$, $\psi_1^j$ satisfies the coupled non-linear 
differential equations, $\left[-\frac 12\partial^2_{x}-\mu^j_{\text{eff}}-
4\gamma|\psi^{j}_{1}|^{2}\right]\psi^{j}_{1}-\kappa\psi^{j'}_{1}= 0$, where $\gamma = c_{1}/c_{0}$ 
and $j=l \ (r)$ when $j' = r \ (l)$. Consequently, within this approximation, the spinor BECs are described 
by two coupled scalar equations. Each scalar equation is characterized by an effective chemical 
potential $\mu^j_{\text{eff}}=\mu^j-(1+2\gamma){n}^j-q$ and an interaction strength equal to 
$-4\gamma>0$. The corresponding stationary solution is readily obtained~\cite{kaurov_06}
\begin{equation} 
\label{eq:sjvsolution}
\psi^{j}_1 = \left[C\text{tanh}(vx) \pm iA\text{sech}(vx)\right]e^{i\alpha},
\end{equation} 
where $C = \sqrt{(2\gamma{n}^j+q)/8\gamma}$, $v$ and $A$ are constants and the $+(-)$ sign 
corresponds to $j=l \ (r)$. This analytical solution allows us to calculate many properties of the system. For 
example, one can directly find the density $n^{j} = (1+\kappa+\gamma)/(1+\gamma)$ in units of 
$n_{R}$ and chemical potential $\mu^{j} = q/2+(\gamma+1)$ in units of $\epsilon_0$.

\begin{figure}
\includegraphics[width=1.0\columnwidth]{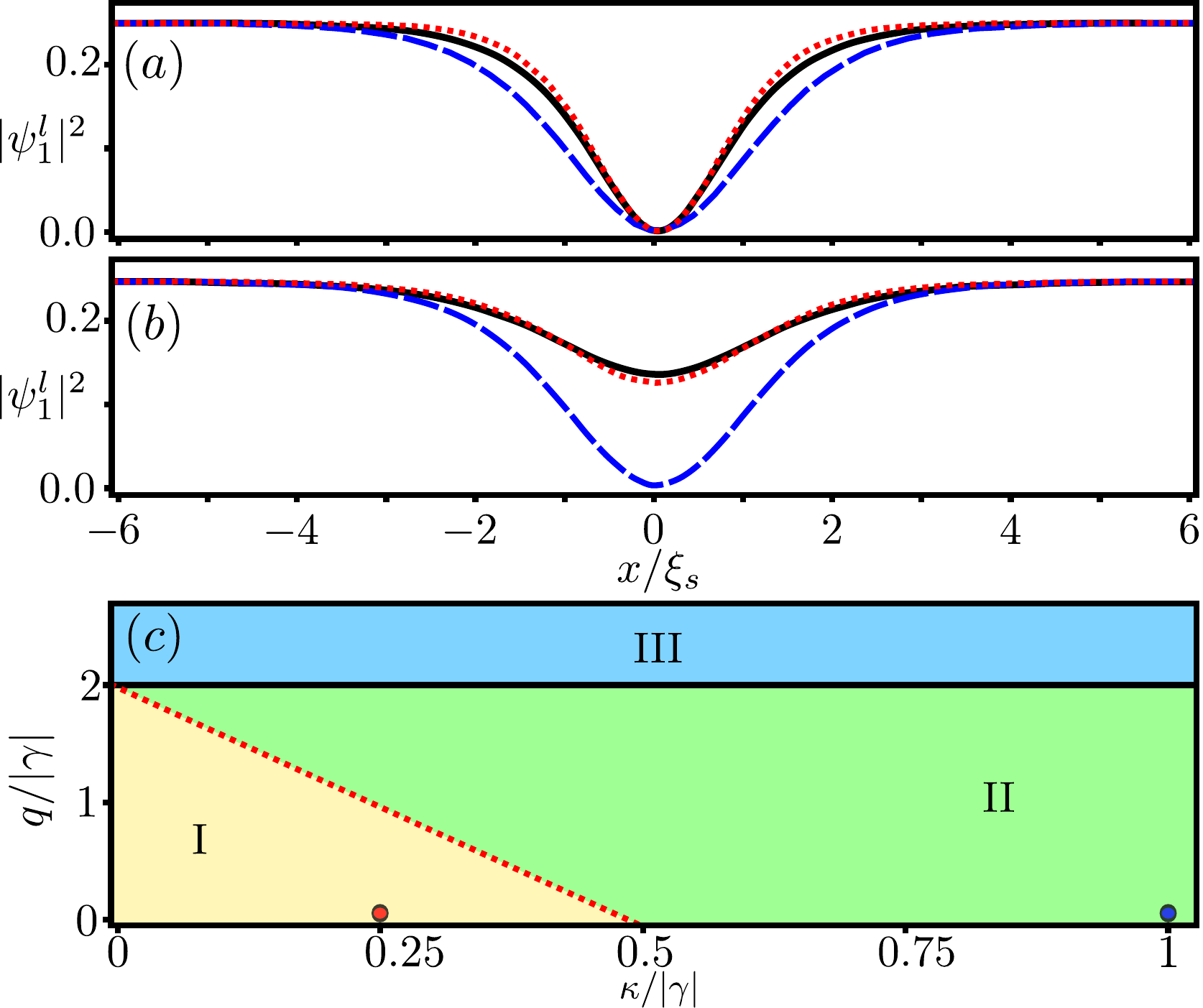}
\caption{(Color online) (a) and (b): $|\psi^l_1(x)|^2$ for a FDW and an 
SJV respectively using the same parameters as in Fig.~\ref{fig1}. The dashed (dotted) curves are calculated 
with (without) the constant atom density approximation and the solid curves show numerical solutions found 
by evolving the GPEs in imaginary time. (c) Phase diagram of the topological excitations. The SJV is dynamically 
stable in region I (yellow). The FDW is unstable in both regions I and II. In region III, the system exhibits a polar 
groundstate phase (see text). Red (blue) circle marks system parameters in region I (II), which are discussed in the text.}
\label{fig2}
\end{figure}        

We now discuss the topological excitations of the system. Depending on the value of $A$, two distinct 
solutions can be obtained from Eq.~(\ref{eq:sjvsolution}). When $A=0$, the solution describes a FDW~\cite{saito_07}, 
which has a characteristic spatial width of $1/v= \sqrt{2}\xi_{s}$, where $\xi_{s} = 1/(2|\gamma|n^j-q)^{\frac{1}{2}}$, is 
the spin healing length. When $A=\sqrt{(2\gamma{n}^j+q+8\kappa)/8\gamma}$, we obtain a totally different topological 
excitation, the \emph{spin Josephson vortex}. The size of an SJV is approximately $1/v=1/\sqrt{4\kappa}$ and, 
hence, controlled by the inter-well tunneling strength, $\kappa$. To distinguish the two distinct 
topological excitations, we calculate their spin texture, characterized by the local spin orientation 
$\phi^{j}(x) = \text{tan}^{-1}(F^j_y/F^j_x)$ and its magnitude 
$|\vec{\bf F}^j| = [(F^j_x)^2 + (F^j_y)^2]^{1/2}$, from Eq.~(\ref{eq:sjvsolution}). The spatial variation of 
the local spin vector along the $x$ axis is shown in Figs.~\ref{fig1}(b,c). In an SJV [Fig.~\ref{fig1}(b)], the 
spin current forms a vortex structure in which the local spin vector rotates between the two spinor BECs 
around a point [blue cross in Fig.~\ref{fig1}(b)] mid-way between them. By contrast, there is no spin 
current associated with the FDW. Instead, the spin vectors in the two atom clouds are locally aligned for 
all $x$ and vanish at $x=0$ [Fig.~\ref{fig1}(c)].
\begin{figure*}
\centering
\includegraphics[width=1.95\columnwidth]{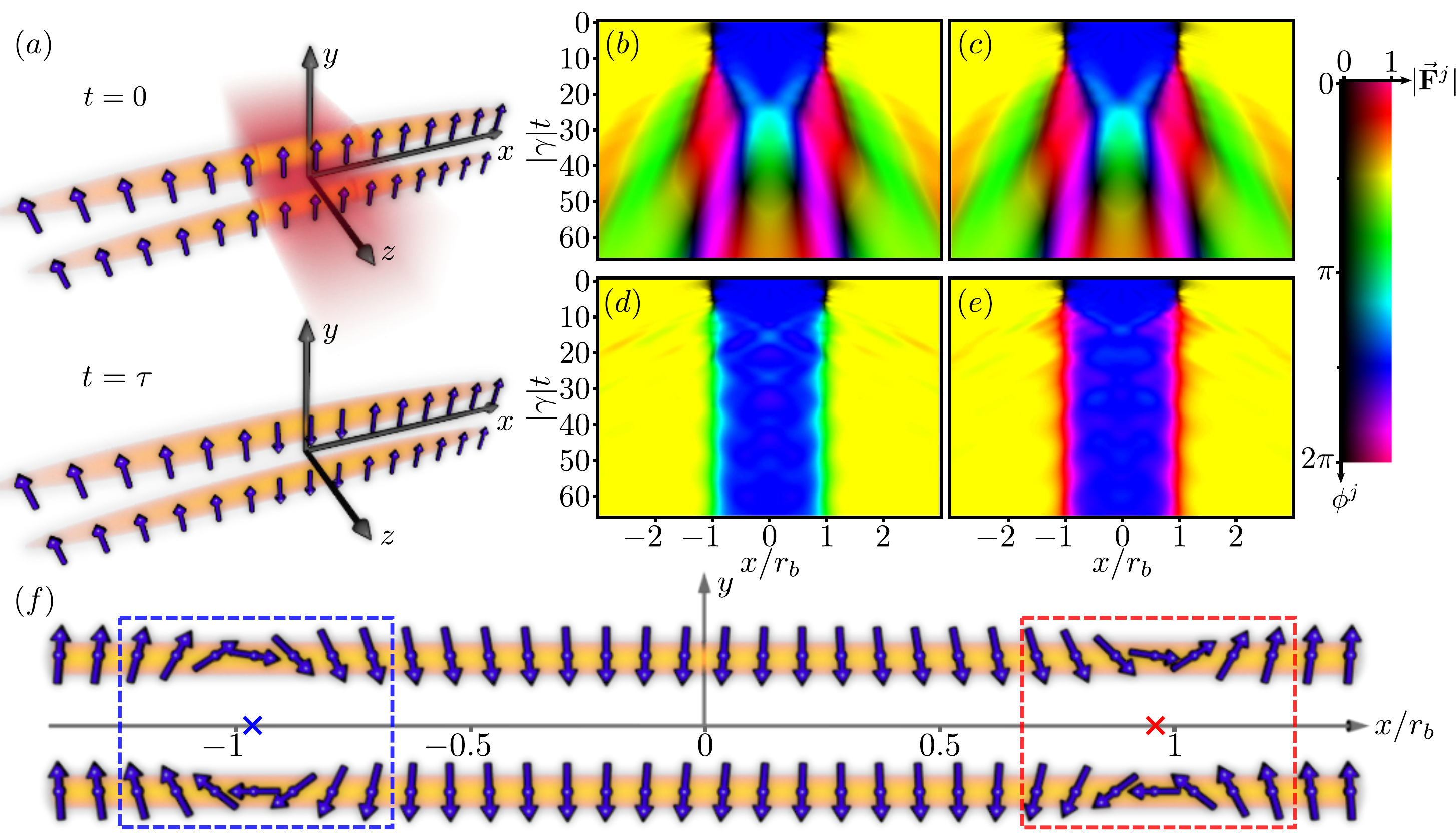}
\caption{(Color online) (a) Schematic diagram showing phase imprinting of the coupled spinor BECs by a 
focused laser beam (red). The phase-imprinting laser is switched on at $t=0$ and only affects atoms in 
the region $-r_b<x<r_b$ spanned by the beam (upper panel). The laser beam is switched off at time 
$\tau  = \pi/(\beta I_0)$ and coherently flips all atomic spins in the region $-r_b<x<r_b$ (lower panel). (b) 
and (c) Color maps showing how the spin vectors $\vec{\mathbf{F}}^{l}$ and $\vec{\mathbf{F}}^{r}$, respectively, 
evolve after the phase imprint when $\kappa = 2\kappa_{c}$. The orientation ($\phi^j$) and magnitude $(|\vec{\mathbf{F}}^j|)$ 
of the local spin vector in the $x-y$ plane are represented, respectively, by the color and brightness of the images (see scale). 
The two black stripes (where $|\vec{\mathbf{F}}^j|=0$) visible for $|\gamma| t < 5$ at $x/r_b=\pm 1$ show the initial 
formation of two FDWs. At later times the stripes vanish, indicating the decay of the FDWs. (d) and (e) are the 
same as (c) and (d), respectively, except that $\kappa = \kappa_{c}/2$. Again, two black stripes 
centered at $x/r_b=\pm 1$ indicate the initial formation of two FDWs for $|\gamma|t<5$. In this case, 
though, the FDW evolves towards a new quasi-static spin texture [green and red stripes in (d) and (e)]. 
At $|\gamma|t=50$, this spin texture corresponds to the local spin vectors shown in (f). Comparing the 
region of (f) within the blue dashed box to Fig. 1(b), we see that an SJV with $\alpha = \pi/2$ (see text) 
has formed, centered at $x/r_b \approx-1$ (blue cross). An anti-SJV with opposite spin vector rotation 
has formed within the red dashed box centered at $x/r_b\approx1$ (red cross). In our simulations, $r_{x} = 1250$ and 
$r_{b} = 100$ (see text).}
\label{fig3}
\end{figure*}

Although the analytical ansatz in Eq.~(\ref{eq:sjvsolution}) is simple, it produces accurate wavefunctions 
when compared with full numerical solutions of the equations of motion, which we obtain by propagating 
the Gross-Pitaevskii equations (GPEs) for the coupled spinor BECs in imaginary time~\cite{kawaguchi06}. 
Figs.~\ref{fig2}(a,b) reveal a small deviation between $\psi^{l}_{1}(x)$ curves obtained analytically 
(dashed curves) and numerically (solid curves) near the center of the SJV and FDW. This deviation is 
caused by the constant density assumption used in the above analytical calculation. To overcome this, 
we now allow a spatially dependent density perturbation, ${n}^j(x) = {n}^{j} + \delta {n}^j(x)$, in the 
ansatz. The resulting values of $\psi^{l}_{1}(x)$ [dotted curves in Figs.~\ref{fig2}(a,b)] agree much 
better with the values obtained numerically (solid curves).

We are now in a position to obtain the phase diagram of the topological excitations from the above 
stationary solutions. Three distinct phase regions are found, which are summarized in Fig.~\ref{fig2}(c). 
Region I (yellow) shows the parameter space where the SJV exists. The FDW exists both in region I and 
region II (green) and a ``polar'' phase in which all atoms are in the $m=0$ spin level occupies region III 
(blue)~\cite{ho_98,stamper-kurn_2012}. We emphasize that the phase diagram can be determined 
completely analytically. For example, the phase boundary between region I and II occurs along the red 
dotted line in Fig.~\ref{fig2}(c), whose equation is 
$\kappa=\kappa_{\text{c}} = (2|\gamma|-q)(1-|\gamma|)/(4-6|\gamma|)$.

To investigate the dynamical stability of the topological excitations, we use an extended Bogoliubov 
theory~\cite{peth_02} in which we evolve a stationary solution, 
${\bf \Psi}_{\text{s}}^j(x)$, to ${\bf \Psi}^j(x,t) = {\bf \Psi}_{\text{s}}^j(x) + \delta{\bf\Psi}^j(x,t)$ at time 
$t$, where $\delta{\bf\Psi}^j(x,t) = {\bf u}^j(x)e^{-i\lambda t} - {\bf v}^j(x)^{*}e^{i\lambda^{*} t}$ is a 
small perturbation. Linearizing these vector equations with respect to ${\bf u}^j$ and ${\bf v}^j$ yields 
an eigenequation with eigenvectors $({\bf u}^l,{\bf v}^l,{\bf u}^r, {\bf v}^r)^T$ and eigenvalues 
$\lambda$. A stable solution requires that $\rm{Im} (\lambda)=0$~\cite{montgomery_2010}. Analysis 
of the eigenvalues reveals that the SJV is dynamically stable in region I. By contrast, the FDW is unstable 
in all regions of Fig.~\ref{fig2}(c).

Guided by this stability analysis, we now explain how to realize the SJV in experiment. First, a FDW is 
created in the spinor BECs by a phase-imprinting method of the type used previously to generate 
topological excitations~\cite{write_09,weibinli08,becker_2008}. Provided the system is in region I of 
Fig.~\ref{fig2}(c) ($\kappa < \kappa_{c}$), the unstable FDW can decay into the stable SJV. 

We now consider the details of the spin-dependent phase-imprinting process. A phase-imprinting laser 
beam propagating along $z$ is switched on at $t=0$. We approximate the intensity profile of the beam 
along the $x$ direction by a square wave of the form $I(x) = I_0\theta(x\pm r_{b})$, where $I_0$ is the 
laser intensity, $\theta$ is the Heavyside function and $2r_b$ determines the width of the laser beam 
along the $x$ direction. Such a shape can be achieved, for example, by reflecting the laser beam from a 
spatial light modulator \cite{becker_2008}. The laser light is circularly polarized ($\sigma_{+})$ to induce 
a linear Zeeman shift through the spin-dependent A.C. Stark energy shift $p(x) = \beta I(x)$, where 
$\beta$ is a constant~\cite{higbie_05}. Applying the beam for a duration $\tau = \pi/(\beta I_0)$ 
coherently flips the atomic spins near the central region of the atom clouds where $|x| < r_{b}$. Such a 
process is shown schematically in Fig.~\ref{fig3}(a). 

As in typical cold atom experiments, we further assume that the spinor atoms are confined along the $x$ 
direction by a shallow harmonic trap. The atom density of the spinor BECs is then given by a 
Thomas-Fermi profile, $n(x)=n_R [1-(x/r_x)^2]$, where $r_x$ is the Thomas-Fermi radius. Initially, the 
two BECs are prepared in the ferromagnetic groundstate, whose wavefunction is 
$\mathbf{\Psi}^{l,r} = \sqrt{{n}(x)}[-1/2,i/\sqrt{2},1/2]$ when $q=0$, in which the spin vectors point 
along the $y$ direction~\cite{ho_98}. After applying the phase-imprinting laser, we determine the 
dynamics by evolving the coupled GPE for the spinor BECs. We include dissipation in our numerical 
simulations following the methods in~\cite{saito_06}. 

Let us now analyze the dynamics of the atom clouds after the laser illumination. We first consider 
the evolution of a system with parameters located in region II of Fig.~\ref{fig2}(c). Specifically, we 
choose $q/|\gamma|=0$ and $\kappa/|\gamma| = 2\kappa_{c}/|\gamma|$ [marked by the blue circle 
in Fig.~\ref{fig2}(c)], for which the evolution of the local spin vectors in the left and right atom clouds 
is shown in Fig.~\ref{fig3}(b) and (c) respectively. For short times ($|\gamma|t < 10$), the system reacts 
to the laser by quickly forming two FDWs at the edges of the phase-imprinting region 
($x/r_{b} = \pm 1$). These appear in Figs.~\ref{fig3}(b,c) as dark stripes, where the magnitude of the 
atomic spin vectors $|\vec{\mathbf{F}}^j|=0$. The stripes separate two bright yellow regions, 
where the atom spins point along the $y$ axis, from a bright blue region where the atom spins point 
along the $-y$ direction. Due to their instability, the FDWs decay into spreading spin textures when 
$|\gamma|t >10$. This decay appears in Figs.~\ref{fig3}(b,c) as muti-colored bands, which emerge from 
the black stripes and spread outwards with increasing $t$.

The evolution of the spin texture differs markedly when the system parameters are prepared in region I 
of Fig.~\ref{fig2}(c).  To illustrate this, we choose $q/|\gamma|=0$ and 
$\kappa/|\gamma| = \kappa_{c}/(2|\gamma|)$ [marked by the red circle in Fig.~\ref{fig2}(c)], for which 
the evolution of the local spin vectors is shown in Figs.~\ref{fig3}(d,e). Comparison of 
Figs.~\ref{fig3}(d,e) with Figs.~\ref{fig3}(b,c) shows that in both cases the system initially 
(for $|\gamma|t < 10$) forms two FDWs (black stripes at $x/r_{b} = \pm 1$). However, instead of 
decaying into a spreading spin texture, in Figs.~\ref{fig3}(d,e), the FDW evolves towards a different 
quasi-static spin pattern, which appears as two vertical bright green [Fig.~\ref{fig3}(d)] or bright red 
[Fig.~\ref{fig3}(e)] stripes centered at $x/r_{b} = \pm 1$. To demonstrate that these spin textures 
correspond to SJV formation, in Fig.~\ref{fig3}(f) we show the associated spin vector configurations in 
the two spinor BECs at $|\gamma|t = 50$. Comparison of the spin vectors around the point 
$x/r_{b} =  -1$ (within the blue dashed box) with Fig.~\ref{fig1}(b) clearly shows that an SJV with 
$\alpha=\pi/2$ has formed. Rather less obviously, a so called anti-SJV has formed at $x/r_{b} = 1$
(within the red dashed box). This corresponds to a solution in Eq.~\eqref{eq:sjvsolution} with $\alpha = -\pi/2$.

In practice, we can create the SJV using, for example, $^{87}$Rb BECs. If the total number of atoms is 
$2\times10^{6}$ and $r_{x} (r_{\perp}) = 200\, \mu$m ($2.4\,\mu$m), the characteristic timescale is 
$|\gamma|t_{0} = 16$ ms. The atomic tunneling strength, $\kappa$, can be controlled by changing the 
intensity and/or waist of the laser that creates the double-well trap~\cite{Albiez05}. All of the system 
parameters and procedures required to implement our proposed route to creating SJVs can be attained 
using current experimental setups~\cite{stamper-kurn_2012}. Consequently, we expect that the 
dynamical regime that we have identified will be directly accessible to experimental study. 

In conclusion, we have identified SJVs in spin-1 ferromagnetic BECs trapped in an elongated DW 
potential. We have presented a detailed analysis of the stability and formation of the SJVs. In particular, 
we have shown that the SJV can be created from the decay of a FDW. Our analysis can be extended to 
study topological phases in multi-well optical potentials and for higher atomic spins, which seem certain 
to reveal further exotic spin textures.

\begin{acknowledgments}
This work is funded by EPSRC. WL acknowledges funding through an EU Marie Curie Fellowship.
\end{acknowledgments}


\begin{thebibliography}{39}
\expandafter\ifx\csname natexlab\endcsname\relax\def\natexlab#1{#1}\fi
\expandafter\ifx\csname bibnamefont\endcsname\relax
  \def\bibnamefont#1{#1}\fi
\expandafter\ifx\csname bibfnamefont\endcsname\relax
  \def\bibfnamefont#1{#1}\fi
\expandafter\ifx\csname citenamefont\endcsname\relax
  \def\citenamefont#1{#1}\fi
\expandafter\ifx\csname url\endcsname\relax
  \def\url#1{\texttt{#1}}\fi
\expandafter\ifx\csname urlprefix\endcsname\relax\def\urlprefix{URL }\fi
\providecommand{\bibinfo}[2]{#2}
\providecommand{\eprint}[2][]{\url{#2}}

\bibitem{ho_98}
\bibfnamefont{T.-L.~Ho},
Phys. Rev. Lett. 
\textbf{81}, 
742 
(1998).

\bibitem{ohmi_1998}
\bibnamefont{T.~Ohmi and K.~Machida}, 
J. Phys. Soc. Jpn. \textbf{67}, 1822 (1998)

\bibitem{schmal_04}
\bibfnamefont{H.~Schmaljohann,
 M.~Erhard, J.~Kronjager, M.~Kottke, S.~van Staa, L.~Cacciapuoti, J.~J.~Arlt, K.~Bongs, and K.~Sengstock}, Phys. Rev. Lett. \textbf{92}, 040402 (2004).

\bibitem{bigelow_05}
\bibfnamefont{N.~Bigelow},
Nat. Phys \textbf{1}, 89 (2005).

\bibitem{chang_05}
\bibfnamefont{M.-S. Chang, Q.~Qishu, W.~Zhang, L.~You, and M.~S. Chapman},
Nat Phys. \textbf{1},
111116 (2005).

\bibitem{sadler_06}
\bibfnamefont{L.~E. Sadler, J.~M. Higbie,S.~R. Leslie, M.~Vengalattore, and D.~M. Stamper-Kurn},
Nature 443, 312 (2006).

\bibitem{mizushima_2002}
\bibnamefont{T.~Mizushima, K.~Machida and T.~Kita}
Phys. Rev. Lett. 89, 030401 (2002)

\bibitem{saito_06}
\bibnamefont{H.~Saito, Y.~Kawaguchi, and M.~Ueda}, Phys. Rev. Lett. 96, 065302 (2006).

\bibitem{write_09}
 \bibnamefont{K.~C. Wright, L.~S. Leslie, A.~Hansen, and N.~P. Bigelow},
 Phys. Rev. Lett. \textbf{102}, 030405 (2009).

\bibitem{kawag_08}
\bibnamefont{Y.~Kawaguchi, M.~Nitta, and M.~Ueda}, Phys. Rev. Lett. \textbf{100}, 180403 (2008).

\bibitem{usama_01}
\bibfnamefont{A.~K.Usama, and H.~Stoof},
 Nature \textbf{411}, 918 (2001).

\bibitem{leslie09}
\bibfnamefont{L.~S. Leslie, A.~Hansen, K.~C. Wright, B.~M. Deutsch, and N.~P. Bigelow}, 
Phys. Rev. Lett. \textbf{103}, 250401 (2009).

\bibitem{choi12}
\bibnamefont{J.-y. Choi, W.~J. Kwon, and Y.-i. Shin},
Phys. Rev. Lett. \textbf{108}, 035301 (2012).

\bibitem{pu_01}
\bibfnamefont{H.~Pu,W.~Zhang, and P. Meystre},
Phys. Rev. Lett. \textbf{87}, 140405 (2001).

\bibitem{demler_02}
\bibfnamefont{E.~Demler, and F.~Zhou},
Phys. Rev. Lett. \textbf{88}, 163001 (2002).

\bibitem{pu_02}
\bibfnamefont{H.~Pu, W.~Zhang, and P.~Meystre},
Phys. Rev. Lett. \textbf{89}, 090401 (2002).
	
\bibitem{yip_03}
\bibnamefont{S.~K. Yip},
Phys. Rev. Lett. \textbf{90}, 250402 (2003).

\bibitem{rizzi_05}
\bibfnamefont{M.~Rizzi, D.~Rossini,G.~De~Chiara, S.~Montangero and R.~Fazio},
Phys. Rev. Lett. \textbf{95}, 240404 (2005).

\bibitem{song_07}
\bibfnamefont{J.~L. Song, G.~W. Semenoff, and F.~Zhou},
Phys. Rev. Lett. \textbf{98}, 160408 (2007).

\bibitem{batrouni_09}
\bibnamefont{G.~G. Batrouni, V.~G. Rousseau, and R.~T. Scalettar}, 
Phys. Rev. Lett. \textbf{102}, 140402 (2009). 

\bibitem{rodriguez_11}
\bibfnamefont{K.~Rodriguez,A.~Arguelles, A.~K. Kolezhuk, L.~Santos, and T. ~Vekua},
Phys. Rev. Lett. \textbf{106}, 105302 (2011).

\bibitem{bloch2005}
\bibfnamefont{I.~Bloch},
Nature Physics \textbf{1}, 23 (2005).

\bibitem{Albiez05}
\bibnamefont{M.~Albiez,
  R.~Gati,
  J.~F\"olling,
  S.~Hunsmann,
  M.~Cristiani,
  and M.~K.
  Oberthaler}, Phys. Rev. Lett.
  \textbf{95}, 010402 (2005).

\bibitem{milburn_97}
\bibfnamefont{G.~J. Milburn, J.~Corney, E.~M. Wright, and D.~F. Walls},
Phys. Rev. A \textbf{55}, 4318 (1997).

\bibitem{you_05}
\bibfnamefont{O.~E. M\"ustecapl\i o \u glu, M.~Zhang, and L.~You},
Phys. Rev. A \textbf{71}, 053616 (2005).


\bibitem{you_07}
\bibnamefont{O.~E. M\"ustecapl\i o\u glu,
  W.~Zhang, and
  L.~You},
  Phys. Rev. A \textbf{75},
  023605 (2007).

\bibitem{julia_09}
\bibnamefont{B.~Julia-Diaz,
  M.~Mele-Messeguer,
  M.~Guilleumas,
  and A.~Polls},
  Phys. Rev. A \textbf{80},
  043622 (2009).

\bibitem{messeguer_11}
\bibnamefont{M.~Melé-Messeguer,
  B.~Juliá-Díaz,
  M.~Guilleumas,
  A.~Polls, and
  A.~Sanpera},
  New J. Phys. \textbf{13},
  033012 (2011).

\bibitem{wzhang_2012}
\bibnamefont{Dan-Wei Zhang, Li-Bin Fu, Z. D. Wang, and Shi-Liang Zhu}, 
Phys. Rev. A \textbf{85},  043609 (2012)

\bibitem{saito_052}
\bibnamefont{H.~Saito and
  M.~Ueda},
  Phys. Rev. A \textbf{72},
  023610 (2005).

\bibitem{saito_07}
\bibnamefont{H.~Saito,
  Y.~Kawaguchi,
  and M.~Ueda},
  Phys. Rev. A \textbf{75},
  013621 (2007).

\bibitem{zhang05}
\bibnamefont{W.~Zhang and
  L.~You},
  Phys. Rev. A \textbf{71},
  025603 (2005).

\bibitem{kaurov_06}
\bibnamefont{V.~M. Kaurov and
  A.~B. Kuklov},
  Phys. Rev. A \textbf{73},
  013627 (2006).

\bibitem{kawaguchi06}
\bibnamefont{Y.~Kawaguchi,
  H.~Saito, and
  M.~Ueda},
  Phys. Rev. Lett. \textbf{97},
  130404 (2006).

\bibitem{stamper-kurn_2012}
\bibnamefont{D.~M. Stamper-Kurn
  and M.~Ueda},
  arXiv:1205.1888  (2012).

\bibitem{peth_02}
\bibnamefont{C.~Pethick and H.~Smith},
  \emph{Bose-Einstein condensation in dilute gases}
  (Cambridge University Press, 2002).

\bibitem{montgomery_2010}
\bibnamefont{T.~W.~A. Montgomery,
  R.~G. Scott,
  I.~Lesanovsky,
  and T.~M.
  Fromhold}, Phys. Rev. A
  \textbf{81}, 063611
  (2010).

\bibitem{weibinli08}
\bibnamefont{W.~Li,
  M.~Haque, and
  S.~Komineas},
  Phys. Rev. A \textbf{77},
  053610 (2008).

\bibitem{becker_2008}
 \bibnamefont{C.~Becker, S.~Stellmer,P.~Soltan-Panahi,S.~Dorscher, M.~Baumert,
  E.~Richter,
  J.~Kronjager,
  K.~Bongs, and
  K.~Sengstock},
  Nature Physics \textbf{4},
  496 (2008).

\bibitem{higbie_05}
\bibnamefont{J.~M. Higbie,
  L.~E. Sadler,
  S.~Inouye,
  A.~P. Chikkatur,
  S.~R. Leslie,
  K.~L. Moore,
  V.~Savalli, and
  D.~M. Stamper-Kurn},
  Phys. Rev. Lett. \textbf{95},
  050401 (2005).

\end{thebibliography}
\end{document}